\documentclass{PoS}
\newcommand{\gcrt}{GCRT\,J1745$-$3009}
\newcommand{\arcmin}{\hbox{$^\prime$}}
\newcommand{\um}{$\mu$m}
\newcommand{\arcsec}{\hbox{$^{\prime\prime}$}}
\newcommand{\lsim}{\mathrel{\hbox{\rlap{\hbox{\lower4pt\hbox{$\sim$}}}\hbox{$<$}}}}

\title{A Bursting Radio Transient in the Direction of the Galactic Center}

\ShortTitle{Bursting Radio Transient}

\author{\speaker{Paul S. Ray}\\
		Space Science Division, Naval Research Laboratory, Washington, DC, USA\\
        E-mail: \email{paul.ray@nrl.navy.mil}}
\author{Scott D. Hyman\\
	Department of Physics and Engineering, Sweet Briar College, Sweet Briar, VA, USA\\
	E-mail: \email{shyman@sbc.edu}}
\author{T. Joseph Lazio, Namir E. Kassim\\
	Remote Sensing Division,	Naval Research Laboratory, Washington, DC, USA\\
	E-mail: \email{lazio@nrl.navy.mil}, \email{namir.kassim@nrl.navy.mil}}
\author{Subhashis Roy\\
ASTRON, Dwingeloo, The Netherlands\\
E-mail: \email{roy@astron.nl}}
\author{David L. Kaplan, Deepto Chakrabarty\\
MIT Kavli Institute, Cambridge, MA, USA\\
E-mail: \email{dlk@space.mit.edu}, \email{deepto@space.mit.edu}}

\abstract{The radio sky is poorly sampled for rapidly varying transients because of
the narrow field-of-view
of most imaging radio telescopes at cm and shorter wavelengths. The
emergence of sensitive long
wavelength observations with intrinsically larger fields-of-view are
changing this situation, as partly illustrated
by our ongoing meter-wavelength monitoring observations and archival studies
of the Galactic Center.
 In this search, we discovered a transient, bursting, radio source in the direction of the Galactic Center, GCRT J1745$-$3009, with extremely unusual properties. Its flux and rapid variability imply a brightness temperature $>10^{12}$ K if it is at a distance $>70$ pc, implying that it is a coherent emitter. I will discuss the discovery of the source and the subsequent re-detections, as well as searches for counterparts at other wavelengths, and several proposed models. }

\FullConference{Bursts, Pulses and Flickering: Wide-field monitoring of the dynamic radio sky June 12-15 2007\\
Kerastari, Tripolis, Greece}

\begin{document}

\section{Introduction}

It is an interesting fact that relatively few transient radio sources are known, particularly when one considers only radio-selected sources. That is, most that are found are triggered by transient phenomena at other wavelengths such as supernova, GRBs, magnetar flares, and microquasar outbursts. In fact, a great deal of radio astronomy makes the tacit assumption that the sky is unchanging. Common examples include sky surveys and multi-configuration imaging.

Why is this the case? Certainly the answer is partly intrinsic.  Many bright radio sources come from extended regions such as supernova remnants and radio lobes that physically cannot vary rapidly. However, most radio observations have been performed and analyzed in a way that is highly insensitive to transient source detection. There are several reasons for this.  Many observations have moved up to high frequencies in search of high angular resolution and since the primary beam width is inversely proportional to frequency, this rapidly reduces the solid angle coverage of an observation. To make matters worse, most observers have a specific target in mind and only image a small field of view, ignoring most of the primary beam. Finally, limited observing time at the major interferometers means that most fields are revisited only rarely resulting in extremely sparse sampling for any transient source searches.

Pushing to low frequencies in search of a naturally larger primary beam
requires sufficient angular resolution to mitigate confusion and provide
good sensitivity. This requires long interferometer baselines (> 5 km), and
has been historically very challenging because ionospheric and 3-D effects
require computationally-intensive algorithms
to enable sensitive wide-field imaging. Only recently has the computational
power and software become
available to enable ionospheric calibration techniques and faceted
wide-field imaging
algorithms to be used routinely \cite{kle+07}. Nevertheless, there are many potential
classes of transient radio
sources \cite{clm04} that could probe some of the most interesting questions in
astrophysics.

This situation contrasts strongly with that in the X-ray band, where the full sky is routinely monitored at a range of timescales.  The most prominent example is the All-Sky Monitor (ASM) on the Rossi X-ray Timing Explorer (RXTE), which has surveyed the full sky many times per day for over a decade.  This has resulted in a great understanding of the population of bursting and transient sources on timescales from hours to years.  Many of these transient X-ray sources involve highly accelerated electrons, often in the form of jets, that also emit in the radio.  In addition there could be other populations of sources that produce radio transients without strong X-ray emission. It seems likely that the paucity of radio transients arises predominately from a lack of sensitive, wide-field observations, a situation that will dramatically change with the advent of many new large-etendue telescopes over the next few years.

\section{GC Monitoring Program}

The Galactic center (GC) is a promising region in which to search for highly variable and transient sources.  The stellar densities are high, and neutron star and black hole binaries appear as (transient or variable) X-ray sources concentrated toward the GC. Previous surveys have been ill-suited for detecting radio transients
toward the GC, however. Typically, they have utilized either single dish instruments, which suffer from confusion in the inner Galaxy, or they have lasted for only a single epoch.

The first two radio transients detected toward the GC were 
A1742$-$28 \cite{dwben76} and the Galactic Center Transient (GCT) \cite{zhaoetal92}. These two transients had similar radio properties, but only the former was associated with an X-ray source. Many radio counterparts to X-ray transients have been detected; e.g., XTE~J1748$-$288 \cite{hrm98a,hrgwm98b} and GRS~1739$-$278 \cite{hrmmr96} have been detected in the \hbox{GC}. 

Recent developments in 3-dimensional imaging techniques \cite{lklh00} have resulted in wide-field images ($\approx 2.5^\circ$) with uniform and high resolution across the field. The discovery \cite{hlkb02} of the radio transient GCRT J1746$-$2757 in our 1998 archival 0.33~GHz observations as well as the detection of the radio counterpart to the X-ray transient XTE J1748-288 provided the impetus for a dedicated search for low frequency radio transients. 

In 2002, we initiated a high resolution, high sensitivity monitoring program of the GC with the VLA at~0.33~GHz in order to detect highly variable and transient radio sources. We extended this program in 2006 to include monitoring observations with the GMRT at 235 MHz. The long-term goal of this program is to constrain the nature of the transient and variable source population(s) based on their individual and group properties. As the GC contains many exotic phenomena not seen elsewhere in the Galaxy, this search may yield the discovery of previously undetected novel classes of radio sources.


\section{Discovery and Characteristics}

In addition to our dedicated monitoring observations, we also re-reduced and analyzed all previous low frequency VLA data taken on the GC region, specifically to look for transients over the full primary beam of the instrument. It was in one of these archival observations that \gcrt\ was discovered.  It was initially detected as a new $\sim 100$ mJy source in a full 6-hour synthesis image.  However, upon breaking up the full observation into short time slices and imaging each separately, it became clear that the source was, in fact, producing a series of $\sim 1$ Jy bursts, each with a duration of $\sim 10$ min, occuring at regular intervals of 77 minutes \cite{hlk+05}.

The rapid variability exhibited in the $\sim$2 min burst decays allowed us to constrain the brightness temperature to be $10^{12}(D/70\mathrm{pc})$K, implying that, unless the source is very nearby, it exceeds the upper limit for incoherent synchrotron radiation and is thus a coherent emitter \cite{hlk+05}.

Subsequently, the source was determined to be a \emph{recurrent} transient when another, somewhat weaker, outburst was detected using the Giant Metrewave Radio Telescope (GMRT) \cite{hlr+06} at 330 MHz.  In this 2003 observation, sparse sampling prevented the detection of other bursts nearby so it is not clear if the 77-min periodicity was still present.

Most recently a very weak single burst was detected in archival data taken at the GMRT in 2004 \cite{hrp+07}, again at 330 MHz. Recent results from analysis of this burst are presented elsewhere in this volume \cite{rhp+07}.

\section{Search for High Energy Counterpart}

Perhaps the most promising way to constrain the potential models for this source would be to discover a counterpart at another wavelength. A detection at X-ray or $\gamma$-ray energies would be particularly powerful at probing the energetics and emission mechanism responsible for the outbursts.  In addition, high energy photons can, like radio waves, penetrate a large amount of intervening matter that is certain to be present if \gcrt\ is at a distance comparable to the Galactic Center.

As described in \cite{hlk+05} there was a serendipitous RXTE/PCA observation pointed 32\arcmin\ from the position of \gcrt\ during the discovery observations.  Unfortunately, the X-ray data did not overlap the actual bursts, and the PCA is not an imaging instrument, so could only place fairly weak constraints on the interburst emission. In addition, all of the RXTE/PCA Galactic bulge monitoring scans were searched for evidence of transient outbursts from this position but no conclusive X-ray counterpart was identified.

Since we know the source is a recurrent transient, it is worth searching for archival detections at that position from any X-ray instruments that monitored the Galactic Center region or the full sky.

A reanalysis of the RXTE/ASM archive (1995--2004) to search for activity at the position of the transient was performed by Ron Remillard (pers. comm.). He finds no evidence for any short timescale (minutes--hours) flares from the source at the $> 100$ mcrab level. His search of the one-day averages reveal a single candidate flare of 26$\pm$3 mcrab on 1996 May 17. Since it is a single event and not correlated with any particular event in the radio it is impossible to exclude the possibility of a statistical fluctuation or an unmodeled systematic error, such as leakage from a flaring source elsewhere in the field. We thus consider this as an upper limit of $\sim 30$ mcrab on the 1-day average flux in the 1.5--12 keV band during the RXTE mission.

A similar analysis was performed by Erik Kuulkers (pers. comm.) on the INTEGRAL Galactic Bulge Monitoring data using the ISGRI 20--60 keV data. 
In addition, Jean in 't Zand (pers. comm.) searched the BeppoSAX WFC data archive for detections of \gcrt,
on time scales of 1 min, 1 hr, 1 day, 0.5 yr and the full 6 years that
the observations span. For the short time scales he looked in two energy
bands: 2--5 and 5--28 keV. The total net exposure is a little over 6 Msec.
The typical sensitivity is 20 mCrab in 1 min to $\sim$1 mCrab for the longest
time scales where the sensitivity is limited by systematic noise.
In both instruments, no detections were made at the position of \gcrt.

\begin{figure}
\includegraphics[width=0.48\textwidth]{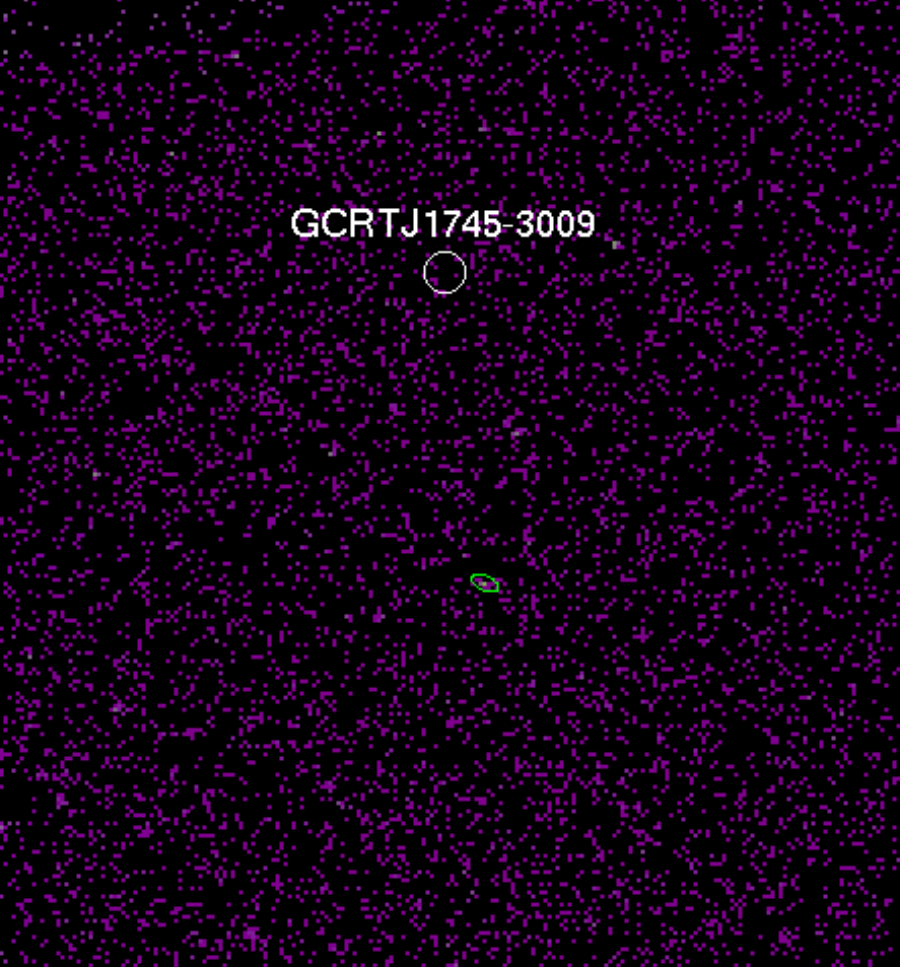}\hfill\includegraphics[width=0.48\textwidth]{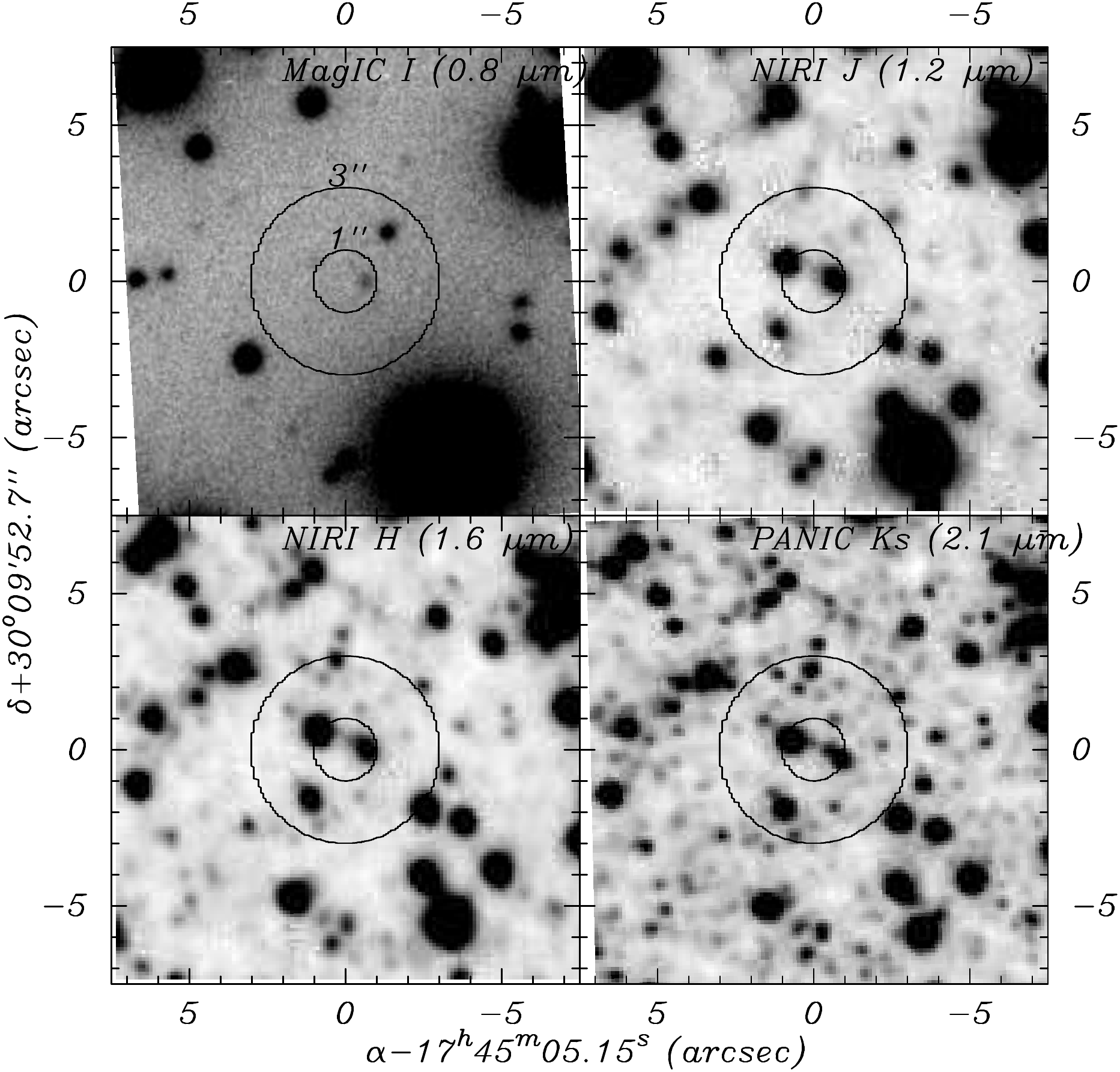}
\caption{(\textit{Left}) Central portion of the Chandra image of the field of \gcrt. A 10 arcsec radius circle indicates the position of \gcrt. The small green ellipse in the lower portion of the image shows the only detected source on this CCD. (\textit{Right}) Images of the field of \gcrt.  We show our best images in $I$
   (top left), $J$ (top right), $H$ (bottom left) and $K_s$ (bottom
   right) bands.  The error circles have radii of $1\arcsec$ and
   $3\arcsec$, corresponding to 1- and 3-$\sigma$ uncertainties. \label{fig-chandra:ir}}
\end{figure}

To search for faint X-ray emission from the source, we obtained a short Director's Discretionary Time observation of \gcrt\ using the ACIS-S instrument (no grating) on the Chandra X-ray Observatory. The 10-ks observation was performed on 2005 May 1. We constructed exposure-corrected images of the full field in two energy ranges (0.3--10 keV and 2--10 keV) and searched each for sources using the celldetect task (see Figure~\ref{fig-chandra:ir}).   Five sources were detected in the 0.3--10 keV image and one source was detected in the 2--10 keV image.  None of these sources were close to the radio position error circle, so we conclude that \gcrt\ remains undetected in X-rays. We have not calculated precise upper limits for the non-detection but the faintest of the detected sources had fluxes of about $4\times10^{-6}$ photons/cm$^2$/s.
Assuming a characteristic X-ray photon energy of 1 keV in the 0.3--10 keV image, this leads to upper limits on the source luminosity of $5.5\times10^{27}$ erg/s at 70 pc and $8.2\times10^{31}$ erg/s at 8.5 kpc. There were no contemporaneous radio observations at the time of the Chandra observation so we have no information on the radio state at that time, but we assume it was in quiescence.

\section{IR Counterpart Search}

Among the many possibilities discussed for the origin of \gcrt,
several involved low-mass stars or substellar objects (brown dwarfs).
The radio emission (both quiescent and flaring) from brown
dwarfs was initially a surprise \cite{bbb+01}, as it is significantly
stronger than one would have expected from the correlation between
radio and X-ray activity seen for other objects. Several late-type
sources have now been detected at GHz frequencies, in many cases
despite the absence of optical or X-ray emission (see \cite{berger06}
for a review).  These objects exhibit a range of nonthermal flaring
behaviors, from strong, narrowband, fully polarized, bursting emission
with frequency drifts \cite{bp05}; to periodic bursts suggestive of
pulsar-like beaming \cite{hbl+07}; to order-of-magnitude variations
spanning years \cite{adh+07}.  In particular, the behavior found by
\cite{hbl+07} for the M9 dwarf TVLM~513$-$46546 is reminiscent of
\gcrt: bright bursts of coherent emission that follow the
2~hr rotation of the star.

We therefore searched for counterparts to \gcrt\ in the near-infrared
(near-IR) and optical bands.  A single brown dwarf in the error circle
of \gcrt\ would be strong evidence that it is the counterpart, even in
the absence of a precise position, as one expects $<1$ brown dwarf
candidate in a $2\arcmin$ field \cite{rkl+99}.
We concentrated on the near-IR for
several reasons: (1) if the source is at the distance of the Galactic
Center, the average $K$-band extinction, $A_K$, at this position is
approximately 2--2.5 magnitudes, corresponding to a visual extinction
of $\geq$20 magnitudes (see e.g., \cite{dsbb03}); (2) if the source
is a more nearby late-type star, it would be intinsically red; and (3)
the field towards the GC is very crowded, and near-IR observations
often have better seeing that bluer bands.  We augmented the
traditional near-IR bands ($JHK_s$) with $I$-band to aid in color
discrimination (see below).

We obtained near-infrared ($JHK_s$ bands, covering wavelengths
1.2--2.1\um) photometric observations of \gcrt\ in the summer of 2005
with both PANIC \cite{mpm+04} on the 6.5-m Magellan~I/Baade telescope
and NIRI \cite{hji+03} on the 8-m Gemini North telescope.  We also
observed \gcrt\ in the $I$-band (0.8\um) in 2006 with MagIC on
Magellan.  The best images (MagIC in $I$ band, NIRI in $J$ and $H$ bands, and PANIC in
$K_s$-band), shown in Figure~\ref{fig-chandra:ir}, have seeing
of $0.4\arcsec$--$0.5\arcsec$ in all four bands.
After standard reduction, we referenced the astrometry and photometry
of the near-IR data to 2MASS \cite{2mass}.  For the $I$-band data, we
registered the astrometry to that of the PANIC $K_s$-band image, and
used observations of the standard fields L113-339 and NGC~6093
\cite{l92,s00} for the photometry.

The position of \gcrt\ is known in the radio frame to $\pm0.7\arcsec$
in each coordinate (1-$\sigma$).  To be conservative and allow for
astrometric frame uncertainties (which should be $\approx 0.2\arcsec$ in
each coordinate), we use a position uncertainty in the near-IR
of $1\arcsec$, and consider objects within 3-$\sigma$ ($3\arcsec$
radius) of the position of \gcrt.

The severe crowding of the field is immediately obvious from the
$K_s$-band image, but so is the severe extinction and the utility of
the other bands.  For instance, approximately 30 objects are visible
within the $3\arcsec$ circle at $K_s$-band, but only 2 at $I$-band!

\begin{figure}
\hbox{\includegraphics[width=0.5\textwidth]{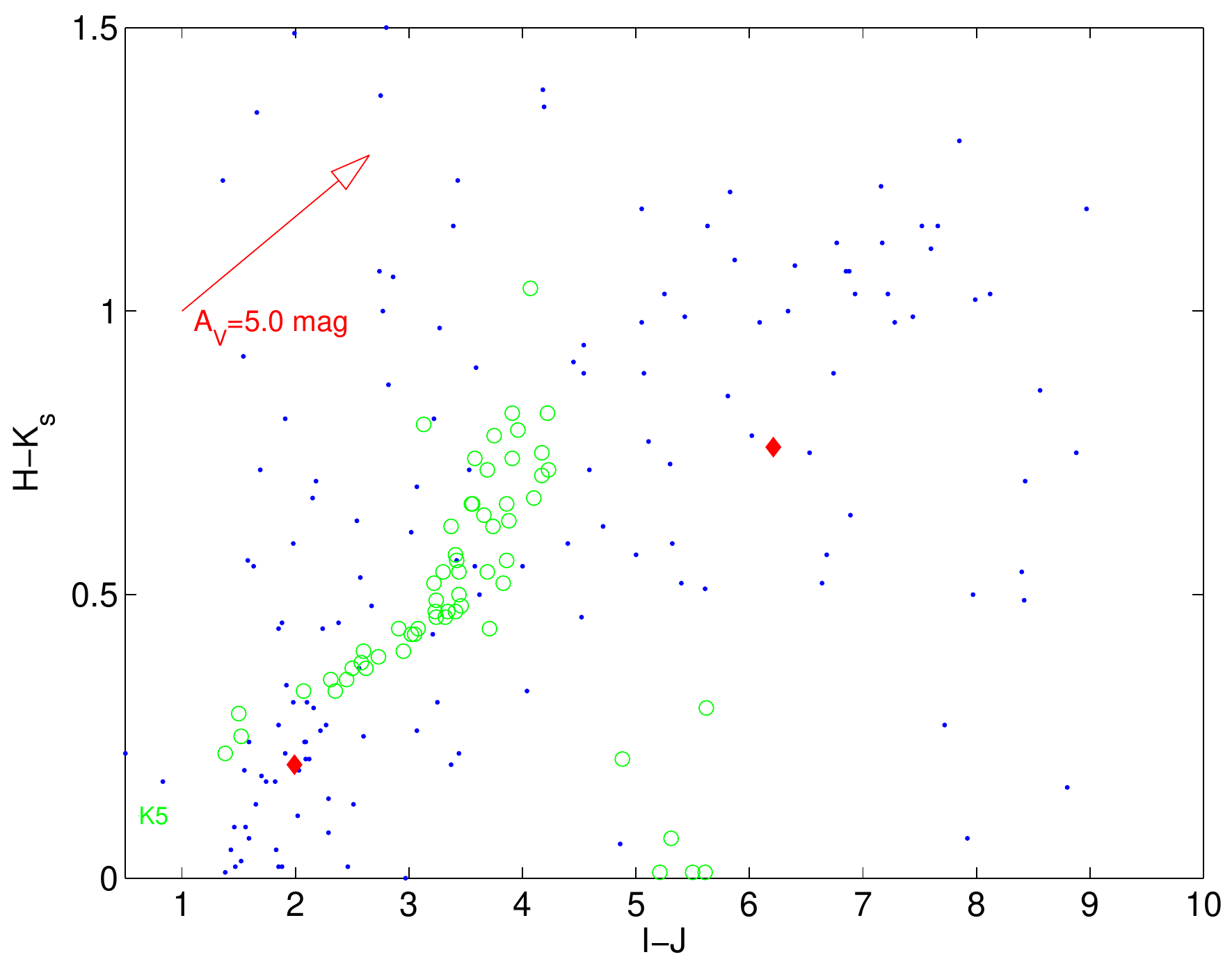} 
\includegraphics[width=0.5\textwidth]{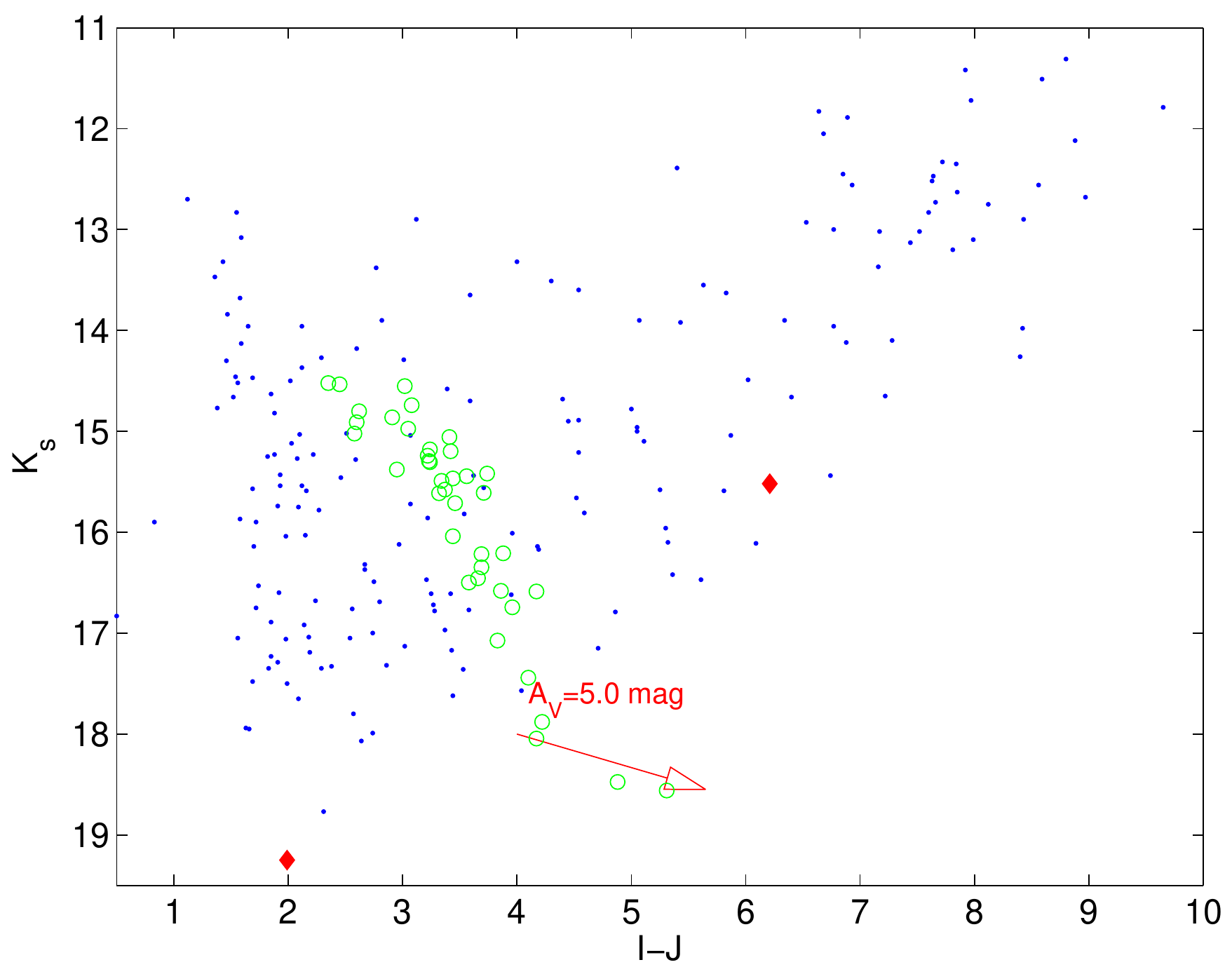}}
\caption{Color-color and color-magnitude diagrams for objects in the
   field of \gcrt.  At left we show $H-K_s$ vs.\ $I-J$ color, while at
   right we show $K_s$ magnitude vs.\ $I-J$ color.  In both panels, the
   blue points are all of the objects detected in the $2\arcmin$
   fields, while the red diamonds are the two objects detected in the
   $I$-band image inside the 3-$\sigma$ error circle.  The green
   circles are the brown dwarf sample from \cite{dhv+02}, ranging in
   spectral type from M8 to T8, and shifted to a distance of 100~pc.
   The arrow shows the effect of 5~mag of extinction.}
\label{fig:cmd}
\end{figure}

Plotting the objects in the field in color-color and color-magnitude
diagrams (Figure~\ref{fig:cmd}), we see a range of spectral types,
with populations of nearby objects at low extinction and more distant,
heavily reddened giants.  Using the $I$-band data helps break some
spectral-type/exinction degeneracies present when only using the
near-IR data.  The depth of the observations allows us to identify
brown dwarfs at distances of $\lsim 100$~pc and stars up to early G
type as far as the GC, as well as \textit{any} evolved star along the
line of sight.  At first glance, neither of the objects detected in
the $I$-band error circle is consistent with being a brown dwarf at a
distance of $<100$~pc: such objects would have $K_s\lsim 18$~mag,
depending on spectral type \cite{dhv+02,tbk03}.  One object appears
to be a late-type star at a distance of $>200$~pc, and the other may
be a more heavily reddened giant, but more detailed analysis is
underway (Kaplan et al. in preparation). We will include proper examination of the stars that are not
detected at $I$-band, to see if the color limits from this help
constrain their nature.  We will search for candidates with anomalous
colors (which may indicate the presence of an accretion disk) and use
the colors of all the sources to measure the extinction across the
field and thus place some constraints on the distances of any
potential IR counterparts.

\section{Pulsar Searches with GBT}

Many of the models for the source involve radio pulsars and thus would be directly confirmable with detection of coherent radio pulsations from the source.  Unfortunately, observations with sensitivity to fast pulses (periods of seconds or shorter) have never been made during periods of known bursting activity.  However, we were able to obtain three observations using the Green Bank Telescope (GBT) during periods of presumed quiescence (see Table~\ref{tab-gbt}).  These observations employed the SPIGOT pulsar backend \cite{kel+05} and are exquisitely sensitive so that even a very faint pulsar in the field of view would be easily detectable.

A relatively coarse analysis of each observation has been performed, including acceleration searches and ``sideband searches'' that increase the sensitivity to pulsars accelerated in a binary orbit, and no strong pulsar candidates were detected.  A finer search, with full sensitivity to periods as short as 1 ms, is planned.

\begin{table}
\caption{Pulsation search observations with the GBT\label{tab-gbt}}
\begin{center}
\begin{tabular}{lcrrrrl}
\hline\hline
Observation & Date & $\nu_0$ & $B$ & $\Delta t$ & $T_\mathrm{obs}$ & Comments \\
               &  & (MHz)       &    (MHz)  & ($\mu$s)    & (hr)   & \\
\hline
4C59-S & 2004-12-30 & 1850 & 600 & 81.92 & 2.6 & Some instrumental issues \\ 
5B9-S & 2005-12-08 & 1850 & 600 & 81.92 & 2.6 &  \\
5B9-P & 2005-12-08 & 350 & 50 & 81.92 & 2.6 &  \\
\hline
\end{tabular}
\end{center}
\end{table}

\section{What is it?}

With no conclusive detections of counterparts at any other wavelength, there is still great uncertainty about the physical nature of the source. 

Given its likely coherent nature, many of the proposed models involved radio pulsars, which are well known coherent emitters clearly capable of producing extremely high brightness temperatures.  The difficulties in modeling involve explaining the variety of timescale exhibited by the source: it is transient on a timescale of months to years, the activity periods last hours to a few weeks, the bursts recur at a period of 77 minutes, and the bursts last less than 10 minutes.  Suggestions have included a ``nulling'' or periodically active pulsar \cite{kp05}, a precessing radio pulsar \cite{zx06}, or a double neutron star binary pulsar \cite{tpt05}. A more exotic suggestion is that of a transient white dwarf pulsar \cite{zg05}. These possibilities are most likely to be confirmed by the detection of pulsations from the source. This may require catching it in its bursting state using a telescope that is sensitive to pulsations but this will be rather difficult, or require substantial good fortune, since the bursting intervals are rather short and occur at unpredictable times.

However, the very recent discovery of periodic bursts of coherent radio emission from an ultracool dwarf (TVLM 513--46546) \cite{hbl+07} have led many to favor a nearby brown dwarf as the counterpart for \gcrt.  Continued work on observations at other wavelengths, particularly the infrared, should allow this model to be confirmed or refuted in the near future.  Since the radio bursts are so bright, one would expect the GCRT to be at a distance closer than TVLM 513--46546, which is at only 10.5 pc \cite{lag+01}.  This is difficult to reconcile with the IR observations that appear to exclude a low mass dwarf out to $\sim$100 pc.

Of course, the possibility still remains that \gcrt\ is an example of a new class of coherent emitter that does not fall into one of these categories.  None of the currently proposed models naturally explain all of the observations.  There are a couple of other observations, each of limited statistical significance, might suggest other avenues to pursue: (1) the source is within the error box of a $>$100 MeV gamma-ray source detected by EGRET (3EG J1744$-$301), and (2) it is also located just outside the shell of SNR 359.1$-$00.5. Because of the crowded nature of the field near the Galactic center, neither of these are strong evidence of an association, but at least the gamma-ray situation will greatly improve with the upcoming launch of GLAST.

\section{Prospects for Future Instruments}\label{sec:prs.future}

A key limitation of our current monitoring program is the relatively
low \emph{survey speed figures of merit} of the VLA and GMRT
\cite{c07}.  The survey speed figure of merit is proportional to
$\Omega A^2$, where $\Omega$ is the field of view of a telescope
and~$A$ is its collecting area.  (See Cordes~\cite{c07} for a more
complete discussion, including additional factors relevant to
transient searching.)  Although both instruments are exceptionally
sensitive (large $A$), they achieve this sensitivity by virtue of
their relatively large individual antennas (25~m and~45~m,
respectively), which result in limited fields of view ($\Omega \approx
1.3 \times 10^{-4}$~sr and $4 \times 10^{-5}$~sr, respectively).
Indeed, despite its being
relatively larger than at more commonly used higher frequencies, the
nevertheless restricted field of view of the 330 MHz VLA was one of our
motivations for targetting the Galactic center in our initial monitoring program.

A straightforward consequence of these small survey speed figures of
merit is that an association between these transients and the Galactic
center remains only probabilistic: While the line of sight toward the
Galactic center has the highest stellar density in the Galaxy, we
cannot exclude the possiblity that objects similar to
GCRT~J1745$-$3009 appear elsewhere on the sky.  Indeed, a brown dwarf
or flare star are among the possible explanations for this object, in
which case we would expect the distribution of such transients to be
nearly uniform on the sky.  

Conducting a survey for transients like GCRT~J1745$-$3009 (or
transients in general) utilizing telescopes like the VLA or GMRT is
not practical.  A useful approach, particularly in cases for which the
luminosity function is not known (as in this case), is to trade
sensitivity ($A$) for field of view ($\Omega$).  Provided of course
that sufficient sensitivity exists to detect GCRT~J1745$-$3009 or
objects like it, making $\Omega$ as large as possible maximizes the
probability of finding similar objects.

Several telescopes under construction, many of them described
elsewhere in this conference (e.g., Long Wavelength Array (LWA), Allen
Telescope Array (ATA), Murchison Widefield Array (MWA), Low Frequency Array (LOFAR)), will
provide much larger survey speed figures of merit; other telescopes
being designed with relatively large survey speed figures of merit
include the Australian SKA Pathfinder\footnote{
\texttt{http://www.atnf.csiro.au/projects/askap/}
} and the Karoo Array Telescope\footnote{
\texttt{http://www.kat.ac.za/}
}.  At a minimum, the survey speed figures of merit of these
instruments are large enough that repeated large-area surveys
(potentially a full hemisphere) are possible.  At best, some of the
telescopes (\hbox{LWA}, \hbox{MWA}) offer the promise of observing
multiple fields of view simultaneously.  Such ``multi-beaming'' would
allow the Galactic center (or other fields such as globular clusters
or nearby galaxies) to be monitored on a frequent basis for future
outbursts from GCRT~J1745$-$3009 or similar objects\footnote{
While LOFAR is likely to offer multi-beaming as well, it is too far
north to observe the Galactic center effectively.
}.  Of course, the spectral characteristics of GCRT~J1745$-$3009 and
similar objects may mean that not all of these telescopes will be
equally favorable for observational programs.  Nonetheless, the next
several years are likely to see rapid progress in recovering
GCRT~J1745$-$3009 or discovering its brethern.

\section*{Acknowledgements}

Basic research in radio and X-ray astronomy at the Naval Research Laboratory is supported by NRL/ONR. S.D.H. is supported by funding from Research Corporation and SAO Chandra grants GO6-7135F and GO6-7038B.

\end{document}